\documentclass[11pt]{article}
\usepackage{enumitem}
\newlist{examples}{enumerate}{1}
\setlist[examples]{label=(\thetable.\arabic*)}
\usepackage{latexsym}
\usepackage{amssymb}
\usepackage{amsmath}
\numberwithin{equation}{section}
\usepackage{cite}

\RequirePackage{amsthm}
\RequirePackage{amscd}
\RequirePackage{epsfig}
\RequirePackage{graphics}
\RequirePackage{ifthen}
\RequirePackage{varioref}
\usepackage{color}

\newcommand{\beq}{\begin{equation}}
\newcommand{\eeq}{\end{equation}}
\newcommand{\bea}{\begin{eqnarray}}
\newcommand{\eea}{\end{eqnarray}}
\newcommand{\barr}[1]{\begin{array}}
\newcommand{\earr}{\end{array}}
\newtheorem{theorem}{Theorem}[section]
\newtheorem{definition}{Definition}[section]

\newcommand{\bdf}{\begin{definition}}
\newcommand{\edf}{\end{definition}}
\newcommand{\bth}{\begin{theorem}}
\newcommand{\enth}{\end{theorem}}

\def\ri{{\rm{i}}}
\def\ep{\varepsilon}
\newcounter{abc}

\def\c+{\rlap{\ \raisebox{.2ex}{\scriptsize+}}\supset}

\begin{document}

\title{On the integrability of a new lattice equation found by multiple scale
analysis}
\author{Christian Scimiterna$^1$, Michael Hay$^2$ and Decio Levi$^1$ }
\maketitle

\noindent
{$^1$Dipartimento di Matematica e Fisica, Universit\`a degli Studi Roma Tre and INFN Sezione di Roma Tre, Via della Vasca Navale 84, 00146 Roma, Italy  \\ e-mail:
scimiterna@fis.uniroma3.it,levi@roma3.infn.it\\$^2$INFN Sezione di Roma Tre, Via della Vasca Navale 84, 00146 Roma, Italy \\ e-mail: hay.michael.c@gmail.com}



%

\begin{abstract}
In this paper we discuss the integrability properties of a nonlinear partial difference equation on the square obtained by the multiple scale integrability test from a class of multilinear dispersive equations defined on a four points lattice. 
\end{abstract}



\section{Introduction}
Discrete equations are important in Mathematical Physics and play a double role. 
From one side discrete space time seems to be basic in the description of fundamental phenomena of nature as provided by quantum gravity \cite{gp}. From the other,  discrete equations  provide good numerical schemes for integrating differential equations \cite{brw}.

A classification of integrable partial difference equation has been given by Adler, Bobenko and Suris~\cite{abs1} (ABS) in the simple case of equations defined on four lattice points using the consistency around the cube condition plus  some  discrete symmetry constrains and the tetrahedron property, to be able to get definite results. These results were subsequently generalized by Boll~\cite{bol} by dropping the symmetry constraints. 

In a recent article Heredero, Levi and Scimiterna~\cite{HLS2013}  used the multiple scale expansion to classify integrable real, multilinear, dispersive equations on the  square lattice. The result of this classification gives a set of equations that do not belong to the ABS list.
A few results in this direction are already known in the literature \cite{Viallet,SGR,LY,GY,GH,A} but  the discovery of new examples is always welcomed as it can provide new insight in the classification of integrable equations on the square not belonging to ABS. 

The results presented in \cite{HLS2013} deal with a class of real, autonomous difference equations in the variable $u: \mathbb Z^2
\rightarrow \mathbb R$  defined on a~$\mathbb{Z}^2$ square-lattice
\bea\label{e}
\mathcal Q (u_{n,m},u_{n+ 1,m},u_{n,m + 1}, u_{n + 1,m + 1}; \beta_1, \beta_2,...)=0,
\eea
where the $\beta_i$'s are real constant parameters.  We require that eq. (\ref{e})  be  rewritable  in the form
\begin{equation}\label{a4}
u_{n+1,m+1} = f ^{(1,1)}(u_{n+1,m}, u_{n,m}, u_{n,m+1}) ,
\end{equation}
where 
\bea \label{d1}
\qquad\qquad (f_{u_{n+1,m}}^{(1,1)}, f_{u_{n,m}}^{(1,1)}, f_{u_{n,m+1}}^{(1,1)}) \ne 0 ,
\eea
 and the index $^{(1,1)}$ indicates  that the function $f$ is obtained from (\ref{e}) by solving for the function $u$  at the point $(n+1,m+1)$. Here and in the following, by the lower index $u_{i,j}$ we mean the partial derivative of the function with respect to $u_{i,j}$.
The conditions (\ref{d1}) are necessary conditions to prevent triviality of the equation (\ref{e}).
To be able to propagate through all the $\mathbb Z^2$ plain, we  suppose, as in~\cite{abs1}, that  (\ref{e}) is a multilinear equation in all its variables, i.e. a polynomial equation in its variables with at most fourth order nonlinearity:
\begin{flushleft}
 \bea \label{eq}
 \mathcal Q_{IV}=f_0&+&a_{00}\, u_{00}+ a_{01}\, u_{01} + a_{10}\, u_{10} + a_{11}\, u_{11}+ 
 (\alpha_1{-}\alpha_2) u_{00}\, u_{10} 
 \\ \nonumber
  &+& (\beta_1{-}\beta_2) u_{00}\, u_{01}+ d_1 u_{00}\, u_{11}+ d_2\, u_{01}\, u_{10} + 
 (\beta_1{+}\beta_2)\, u_{10}\, u_{11} \\ \nonumber
 &+&
  (\alpha_1{+}\alpha_2)\, u_{01}\, u_{11}+ 
 (\tau_1{-}\tau_3) u_{00}\, u_{01}\, u_{10}+ (\tau_1{+}\tau_3) u_{00}\, u_{10}\, u_{11} 
\\ \nonumber
 &+& (\tau_2{+}\tau_4) u_{00}\, u_{01}\, u_{11}+ 
 (\tau_2{-}\tau_4)\, u_{10}\, u_{01}\, u_{11} 
+ f_1\, u_{00}\, u_{01}\, u_{10}\, u_{11}=0,
\eea  
\end{flushleft}
where all coefficients have been taken to be real and independent of $n$ and $m$.  
Heredero et. al. considered  the multiple scale expansion around the dispersive solution 
\bea \label{e2}
u_{n,m} = e^{\ri kn} e^{-\ri\omega m},
\eea
of the linearized equation of (\ref{eq}) with a dispersion relation~$\omega=\omega\left(k\right)$
\bea \label{e5+}
\omega_{+}(k) =\arctan\left[\frac{2a_1 a_2 + (a_1^2+a_2^2)\cos(k)}{\left(a_1^2-a_2^2\right)\sin(k)}\right]
\eea
where $f_0=0$, $a_{00}=a_{11}\equiv a_1$, $a_{01}=a_{10}\equiv a_2$, $a_1$ and ~$a_2$ cannot be zero and their ratio cannot be equal to $\pm 1$, to get a nontrivial dispersion relation.
The effective expansion of the dependent variable is
\bea
u_{n,m}= \sum_{\ell=0}^{\infty} \ep^\ell\sum_{\alpha=-\ell-1}^{\ell+1} 
K^{\alpha n} \Omega^{\alpha m}  v_{\ell+1}^{(\alpha)}, 
\label{bas}
\eea
where $v^{(\alpha)}_\ell =v^{(\alpha)}_\ell (n_1, \{m_j\})$ is a bounded,  slowly varying function of its arguments and $v^{(-\alpha)}_{\ell}=\bar v^{(\alpha)}_{\ell}$,
$\bar v_{\ell}$ being  the complex conjugate of $v_{\ell}$ as we are looking at real solutions.
Here $n_1= \ep n$, $m_j = \ep^j m$ $j= 1, 2, \dots$  are the slow-varying lattice 
variables.
 The integrability conditions presented  in \cite{HLS2013} were determined by comparing the various equations obtained through a  multiple scale perturbative expansion with those of an integrable hierarchy.

The results of \cite{HLS2013} are a series of integrability theorems and a table  of equations, invariant under restricted  M\"obius transformations,   that pass the very stringent integrability  conditions obtained by considering the multiple scale expansion up to $\ep^6$ order.
In this way we have obtained the following equations:
\begin{subequations}\label{Quadrelli}
\bea
 &&v_{n,m}+v_{n+1,m+1}+2\left(v_{n+1,m}+v_{n,m+1}\right)+v_{n+1,m}v_{n,m+1}\left(1+\tau\right)+\label{Quadrelli1}\\
 \nonumber &&\quad +\left(v_{n+1,m}v_{n+1,m+1}+v_{n,m}v_{n,m+1}\right)\tau+v_{n,m+1}v_{n+1,m+1}+v_{n,m}v_{n+1,m}+\\
 \nonumber && \quad +v_{n+1,m}v_{n,m+1}\left(v_{n,m}+v_{n+1,m+1}\right)\tau=0; \\
 &&v_{n,m}+v_{n+1,m+1}+\epsilon\left(v_{n+1,m}+v_{n,m+1}\right)+\label{Quadrelli2}\\
 \nonumber &&\quad +\delta\left[\epsilon v_{n+1,m}v_{n,m+1}\left(v_{n,m}+v_{n+1,m+1}\right)+v_{n,m}v_{n+1,m+1}\left(v_{n+1,m}+v_{n,m+1}\right)\right]+\\
 \nonumber && \quad +\tau v_{n,m}v_{n+1,m}v_{n,m+1}v_{n+1,m+1}=0,\, -1<\epsilon<1,\, \epsilon\not=0,\, \delta\doteq\pm 1,\, \tau\geq 0;\\
 &&v_{n,m}+v_{n+1,m+1}+\epsilon\left(v_{n+1,m}+v_{n,m+1}\right)+\label{Quadrelli3}\\
 \nonumber &&\quad +\delta\left[v_{n+1,m}v_{n,m+1}\left(v_{n,m}+v_{n+1,m+1}\right)+\epsilon v_{n,m}v_{n+1,m+1}\left(v_{n+1,m}+v_{n,m+1}\right)\right]+\\
 \nonumber &&\quad +\tau v_{n,m}v_{n+1,m}v_{n,m+1}v_{n+1,m+1}=0,\ \ \ -1<\epsilon<1,\, \epsilon\not=0,\, \delta\doteq\pm 1,\, \tau\geq 0;\\
 &&v_{n,m}+v_{n+1,m+1}+\epsilon\left(v_{n+1,m}+v_{n,m+1}\right)+v_{n,m}v_{n+1,m+1}-v_{n+1,m}v_{n,m+1}+\label{Quadrelli4}\\
 \nonumber &&\quad +\left(1-\frac{1} {\epsilon}\right)\left[v_{n+1,m}v_{n,m+1}\left(v_{n,m}+v_{n+1,m+1}\right)-v_{n,m}v_{n+1,m+1}\left(v_{n+1,m}+v_{n,m+1}\right)\right]+\\
 \nonumber &&\quad +\left(1-\frac{1} {\epsilon^2}\right)v_{n,m}v_{n+1,m}v_{n,m+1}v_{n+1,m+1}=0,\, -1<\epsilon<1,\, \epsilon\not=0,\, \frac{1}{2}.
\eea
\end{subequations}
Here we will consider in detail eq. (\ref{Quadrelli1}) which depends on one free parameter $\tau$. 
If, when $\tau=0$  in (\ref{Quadrelli1}), we apply the transformation $v_{n,m}\doteq\sqrt{3}w_{n,m}-1$, we obtain
\bea
\label{kj1}w_{n,m}w_{n+1,m}+w_{n+1,m}w_{n,m+1}+w_{n,m+1}w_{n+1,m+1}-1=0.
\eea
We look for the generalized symmetries of this equation, which demonstrate its integrability. The existence of 5-point generalized symmetries is an integrability condition for it. 

 If, when $\tau=1$, we apply to (\ref{Quadrelli1}) the transformation $v_{n,m}\doteq -\left(2^{1/3}w_{n,m}+1\right)$, we obtain
\bea
\label{jk2}w_{n+1,m}w_{n,m+1}\left(w_{n,m}+w_{n+1,m+1}\right)+1=0,
\eea
 an integrable equation considered in \cite{MX}, which possesses a $3\times 3$ Lax pair and which is a degeneration of the discrete integrable Tzitzeica equation proposed by Adler in \cite{A}.
 
 Finally, if we choose $\tau\not=0$, $1$  in (\ref{Quadrelli1}), we can apply the  transformation \newline $v_{n,m}\doteq \frac{1-\tau}{\tau}w_{n,m}-1$ and we obtain
\bea
\label{jk3}  && w_{n,m}w_{n+1,m}+w_{n,m+1}w_{n+1,m+1}+\\ \nonumber   &&\qquad \qquad +w_{n+1,m}w_{n,m+1}\left(1+w_{n,m}+w_{n+1,m+1}\right)+\chi=0, 
\eea
 where $\chi\doteq\frac{\left(\tau-3\right)\tau^2}{\left(1-\tau\right)^3}$. 
$$\\$$

In Section \ref{s1} we present a review of the essential notions necessary to derive generalized symmetries for equations on a lattice and in Section \ref{s2} we apply the results of Section \ref{s1} to (\ref{kj1}) and then a sequence of results on (\ref{Quadrelli1}) are obtained. Section \ref{s3} is devoted to some concluding remarks.

\section{Generalized symmetries for quad--graph equations} \label{s1}
As  is well known, the generalized symmetry method allows one to classify integrable nonlinear partial differential equations of a given class and to test a given nonlinear partial differential equation for integrability \cite{mss91,msy87}. In the case of $1+1$ partial differential equations it has been used to develop a computer PC-package DELiA, written in Turbo PASCAL by Bocharov  \cite{delia}. This program can be used  to prove integrability and compute symmetries of given evolutionary partial differential equations. The symmetry approach has also been applied with success to study the integrability of differential difference equations  \cite{asy00,y06} and recently of partial difference equations \cite{ly09,ly10,LY2}. Similar results have been obtained requiring the existence of a recursive operator for the symmetries of the partial difference equation \cite{MWX1,MWX2,MX}.

Given equation (\ref{e}), following ref. \cite{ly09}, we use as an integrability criterion the existence of an autonomous generalized symmetry, which, since (\ref{e}) is autonomous, we can write at the point $n=m=0$ without loss of generality by applying translation invariance. Thus we have:
\bea\label{a4a}
 \quad \frac{d}{dt} u_{0,0} = g_{0,0} = G(u_{\ell,0}, u_{\ell -1,0}, \dots,
u_{\ell',0}, u_{0,k}, u_{0,k-1}, \dots, u_{0,k'}) ,
\eea
with $\ell > 0$, $\ell' < \ell$, $k >0$, $k' < k$ and $t$ denotes the group parameter. The order of the symmetry is defined by  $\ell$ and $k$. Let us notice that as the symmetries of an integrable partial difference equation themselves have to be integrable differential difference equations, they too must satisfy the theorem presented in Section 2.4.1 of \cite{y06}, i.e. they must be symmetric, $\ell'=-\ell$ and $k'=-k$.  If $\ell' \ne -\ell$ and $k' \ne -k$ then the equation  will be linearizable in general and will not have  conservation laws of high order.
It is straightforward to check \cite{LY2,MWX1,GH} that, if
$$G=G(u_{\ell,0}, u_{\ell -1,0}, \dots,
u_{-\ell,0}, u_{0,k}, u_{0,k-1}, \dots, u_{0,-k})$$ is a symmetry of a
quadrilateral equation, then it is a sum of two functions
\bea \label{a4b}
 \qquad G=\Phi(u_{\ell,0}, u_{\ell -1,0}, \dots,
u_{-\ell,0})+\Psi(u_{0,k}, u_{0,k-1}, \dots, u_{0,-k})\,.
\eea
Eq. (\ref{a4a}) is a generalized symmetry of eq. (\ref{a4}) if the following compatibility condition is satisfied:
\bea \nonumber
\frac{d (u_{1,1} - f^{(1,1)})}{dt} \biggl|_{[u_{1,1} = f^{(1,1)}, \, u_{0,0,t}=g_{0,0}]}= 0 .
\eea
 Explicitly the compatibility condition reads:
\begin{equation}\label{a7}
[g_{1,1} - ( g_{1,0} \partial_{u_{1,0}} + g_{0,0} \partial_{u_{0,0}} + g_{0,1} \partial_{u_{0,1}} ) f^{(1,1)}] \biggl|_{u_{1,1} = f^{(1,1)}}=0,
\end{equation}
where $g_{i,j} = T_1^{i} T_2^j g_{0,0}$, and $T_1,T_2$ are the shift operators acting on the first and second indexes, respectively, i.e. $T_1 g_{0,0}=g_{1,0}$ and $T_2 g_{0,0}=g_{0,1}$.
To be able to check the compatibility condition between (\ref{e}) and  (\ref{a4a})  we need to define the set of independent variables in terms of which (\ref{a7}) can be split  into an overdetermined system of independent equations.   In this article we choose the functions
\begin{equation}\label{a8}
 u_{i,0} , u_{0,j}, \, \forall (i,j)
\end{equation}
as {\it independent variables} for any fixed $n$ and $m$. Then, using  (\ref{a4}), all the other functions $u_{i,j}$ can be explicitly written in terms of the independent variables (\ref{a8}). So we require that eq. (\ref{a7}) is satisfied identically for all values of the independent variables.
In \cite{ly09}  the following theorem has been proven:

\begin{theorem}\label{te}
If (\ref{a4}) possesses a generalized symmetry of the form (\ref{a4a}), with $\ell=1$ and $k=1$,  then its solutions  must satisfy the following  conservation laws
\begin{equation}\label{a9}
 (T_1 -1) p_{0,0}^{(\kappa)} = (T_2 -1) q_{0,0}^{(\kappa)} ,
\end{equation}
where
\par
if $ G_{u_{1,0}} \ne 0$, then
\begin{equation}\label{a10}
 p^{(1)}_{0,0} = \log f_{u_{1,0}}^{(1,1)} , \qquad
 q^{(1)}_{0,0} = Q^{(1)}_{0,0}(u_{2,0},u_{1,0},u_{0,0}) ;
\end{equation}
\par
if $ G_{u_{-1,0}} \ne 0$, then
\begin{equation}\label{a11}
 p^{(2)}_{0,0} = \log \frac{f^{(1,1)}_{u_{0,0}}}{f^{(1,1)}_{u_{0,1}}} , \qquad
 q^{(2)}_{0,0} = Q^{(2)}_{0,0} (u_{2,0},u_{1,0},u_{0,0}) ;
\end{equation}
\par
if $ G_{u_{0,1}} \ne 0$, then
\begin{equation}\label{a12}
 q^{(3)}_{0,0} = \log f^{(1,1)}_{u_{0,1}} , \qquad
 p^{(3)}_{0,0} = P^{(3)}_{0,0} (u_{0,2},u_{0,1},u_{0,0}) ;
\end{equation}
\par
if $G_{u_{0,-1}} \ne 0$, then
\begin{equation}\label{a13}
 q^{(4)}_{0,0} = \log \frac{f^{(1,1)}_{u_{0,0}}}{f^{(1,1)}_{u_{1,0}}} , \qquad
 p^{(4)}_{0,0} = P^{(4)}_{0,0} (u_{0,2},u_{0,1},u_{0,0}) .
\end{equation}
\end{theorem}
 Theorem \ref{te} provides four {\it necessary  conditions} for the existence of a five-point  generalized symmetry,  with $\ell$ and $k$ equal unity,  which are written in the form of conservation laws which, in the case of a non-degenerate symmetry (\ref{a4a}), must be all satisfied. In \cite{ly09} and \cite{LY2} the authors where able to prove the integrability of  all the equations included in \cite{abs1}, see also \cite{lpsy}, and also of some new ones, see, for example, \cite{ly091, GY}. 

For the sake of clarity,  we repeat here all  steps necessary to apply the integrality test in the case of $\ell$ and $k$ equal  unity. 

First we consider the integrability conditions (\ref{a9}) with $\kappa=1,2$.
The unknown functions in the right hand side of (\ref{a9}), given by (\ref{a10}, \ref{a11}),  contain the dependent variable $u_{2,1}$ which, from (\ref{a4}), depends on $u_{2,0},\, u_{1,0}, \, u_{1,1}$ and thus it is not immediately expressed in terms of independent variables, but gives rise to extremely complicated functional expressions of the independent variables. We can avoid this problem by applying the operators $T_1^{-1}$ and $T_2^{-1}$ to  (\ref{a9}). In this case (\ref{a9}) is replaced by: 
\begin{equation}\label{a16}
\begin{array}{l}
 p^{(\kappa)}_{0,0} - p^{(\kappa)}_{-1,0} = 
 Q^{(\kappa)}_{-1,1}(u_{1,1},u_{0,1},u_{-1,1}) - Q^{(\kappa)}_{-1,0}(u_{1,0},u_{0,0},u_{-1,0}) ,
\end{array}
\end{equation}
\begin{equation}\label{a17}
\begin{array}{l}
 p^{(\kappa)}_{0,-1} - p^{(\kappa)}_{-1,-1} = 
 Q^{(\kappa)}_{-1,0}(u_{1,0},u_{0,0},u_{-1,0}) - Q^{(\kappa)}_{-1,-1}(u_{1,-1},u_{0,-1},u_{-1,-1}) .
\end{array}
\end{equation}
Here $p^{(\kappa)}_{i,j}$ are known functions expressed in term of (\ref{a4}). The functions $Q^{(\kappa)}_{0,0}$ are unknown, and ($Q^{(\kappa)}_{-1,1}$, $Q^{(\kappa)}_{-1,-1}$) contain the dependent variables $u_{1,1}$, $ u_{-1,1}$, $u_{1,-1}$, $ u_{-1,-1}$ which are expressed in terms of the independent functions through (\ref{e}). Our aim is to derive from (\ref{a16}, \ref{a17}) a set of equations for the unknown function, $Q^{(\kappa)}_{-1,0}$.

To do so let us extract from  (\ref{e}) three further expressions of the form of (\ref{a4}) for the dependent variables contained in  (\ref{a16}, \ref{a17}): 
\bea\label{a18}
  && \qquad \qquad  \qquad \qquad u_{-1,1} = f^{(-1,1)}(u_{-1,0},u_{0,0},u_{0,1}) , \\ \nonumber
  &&u_{1,-1} = f^{(1,-1)}(u_{1,0},u_{0,0},u_{0,-1}) , \;
 u_{-1,-1} = f^{(-1,-1)}(u_{-1,0},u_{0,0},u_{0,-1}) .
\eea
All functions $f^{(i,j)}$ have a nontrivial dependence on the independent variables, as it is the case with $f^{(1,1)}$. 

Let us introduce the two differential operators \cite{hydon,RasinHydon,ly09,LY2}:
\begin{equation}\label{a19}
A = \partial_{u_{0,0}} - \frac{f^{(1,1)}_{u_{0,0}}}{f^{(1,1)}_{u_{1,0}}} \partial_{u_{1,0}} -
\frac{f^{(-1,1)}_{u_{0,0}}}{f^{(-1,1)}_{u_{-1,0}}} \partial_{u_{-1,0}} ,
\end{equation}
\begin{equation}\label{a20}
B = \partial_{u_{0,0}} - \frac{f^{(1,-1)}_{u_{0,0}}}{f^{(1,-1)}_{u_{1,0}}} \partial_{u_{1,0}} -
\frac{f^{(-1,-1)}_{u_{0,0}}}{f^{(-1,-1)}_{u_{-1,0}}} \partial_{u_{-1,0}} ,
\end{equation}
 chosen in such a way to annihilate the functions $Q^{(\kappa)}_{-1,1}$ and $Q^{(\kappa)}_{-1,-1}$,  namely
 \newline $A Q^{(\kappa)}_{-1,1} = 0$, $B Q^{(\kappa)}_{-1,-1} = 0$. Applying $A$ to (\ref{a16}) and $B$ to (\ref{a17}), we obtain two equations for the unknown $Q^{(\kappa)}_{-1,0}$:
\begin{equation}\label{a21}
A Q^{(\kappa)}_{-1,0} = r^{(\kappa,1)} , \quad B Q^{(\kappa)}_{-1,0} = r^{(\kappa,2)} ,
\end{equation}
where $r^{(\kappa,1)}$, $r^{(\kappa,2)}$ are some explicitly known functions of (\ref{a4}). Considering  the standard commutator of $A$ and $B$, $ [A,B] = AB - BA$, we can add a further  equation
\bea \label {a21a}
 [A,B] Q^{(\kappa)}_{-1,0} = r^{(\kappa,3)}.
 \eea

 Eqs. (\ref{a21}, \ref{a21a}) represent a linear partial differential system of three equations for the three derivatives of the unknown functions $q^{(\kappa)}_{-1,0}=Q^{(\kappa)}_{-1,0}(u_{1,0}, u_{0,0}, u_{-1,0})$. For the three partial derivatives of $q^{(\kappa)}_{-1,0}$, this is  just a linear algebraic system of three equations in three unknowns. When this system is non-degenerate it provides one and only one solution for the three derivatives of $q^{(\kappa)}_{-1,0}$.
In these cases we can find  the partial derivatives of $q^{(\kappa)}_{0,0}$ uniquely. Then we can check the consistency of the partial derivatives and, if satisfied, find $q^{(\kappa)}_{0,0}$ up to an arbitrary constant. Finally we fix the constant by checking the integrability condition (\ref{a9}) with $\kappa=1,2$ in either of the equivalent forms (\ref{a16}) or (\ref{a17}).

The non-degeneracy of the system (\ref{a21}, \ref{a21a}) depends on  (\ref{a4}) only.
So, if we have checked the non-degeneracy for  $k=1$, we know that
this is also true for  $k=2$ and vice versa. So both functions $q^{(1)}_{n,m}$, $q^{(2)}_{n,m}$ are found in unique way up to a constant of integration.

If the system (\ref{a21}, \ref{a21a}) is degenerate, the functions $q^{(1)}_{n,m}$, $q^{(2)}_{n,m}$ are defined up to some arbitrary functions. In this case checking the integrability conditions (\ref{a9}) may be more difficult.


Let us consider now the conditions (\ref{a9}) with $\kappa=3,4$. We have a  similar situation as when $\kappa=1,2$. By appropriate shifts we rewrite (\ref{a9}) in the two equivalent forms: 
\bea\label{a22}
  q^{(\kappa)}_{1,-1} - q^{(\kappa)}_{0,-1} = 
 \quad P^{(\kappa)}_{1,-1}(u_{1,1},u_{1,0},u_{1,-1}) - P^{(\kappa)}_{0,-1}(u_{0,1},u_{0,0},u_{0,-1}) ,
\eea
\bea\label{a23}
  q^{(\kappa)}_{0,-1} - q^{(\kappa)}_{-1,-1} = 
  \quad P^{(\kappa)}_{0,-1}(u_{0,1},u_{0,0},u_{0,-1}) - P^{(\kappa)}_{-1,-1}(u_{-1,1},u_{-1,0},u_{-1,-1}) .
\eea
We can introduce the operators
\begin{equation}\label{a24}
\hat A = \partial_{u_{0,0}} - \frac{f^{(1,1)}_{u_{0,0}}}{f^{(1,1)}_{u_{0,1}}} \partial_{u_{0,1}} -
\frac{f^{(1,-1)}_{u_{0,0}}}{f^{(1,-1)}_{u_{0,-1}}} \partial_{u_{0,-1}} ,
\end{equation}
\begin{equation}\label{a25}
\hat B = \partial_{u_{0,0}} - \frac{f^{(-1,1)}_{u_{0,0}}}{f^{(-1,1)}_{u_{0,1}}} \partial_{u_{0,1}} -
\frac{f^{(-1,-1)}_{u_{0,0}}}{f^{(-1,-1)}_{u_{0,-1}}} \partial_{u_{0,-1}} ,
\end{equation}
such that $\hat A P^{(\kappa)}_{1,-1} = 0$ and $\hat B P^{(\kappa)}_{-1,-1} = 0$. Then we are led to the system
\begin{equation}\label{a26}
\hat A P^{(\kappa)}_{0,-1} = \hat r^{(\kappa,1)} , \quad \hat B P^{(k)}_{0,-1} = \hat r^{(\kappa,2)} ,
\quad [\hat A,\hat B] P^{(\kappa)}_{0,-1} = \hat r^{(\kappa,3)}
\end{equation}
for the function $P^{(\kappa)}_{0,-1}$ depending on $u_{0,1},\,u_{0,0},\,u_{0,-1}$, where $\hat r^{(\kappa,l)}$ are known functions expressed in terms of $f^{(i,j)}$,  whose solution is obtained in the same way as for the system (\ref{a21}, \ref{a21a}). 
\bigskip

After we have solved (\ref{a9}--\ref{a13}) we can construct a generalized symmetry. When both systems (\ref{a21}, \ref{a21a}) and (\ref{a26}) are non-degenerate, we find $\Phi$ and $\Psi$, given by (\ref{a4b}), up to at most four arbitrary constants which are specified using the compatibility condition (\ref{a7}).


In the general case when $k$ and $\ell$ are greater than one,  the situation is more complicated. Theorem 2 in \cite{ly09} provides  some conditions which, however,  turn out to be difficult to use for testing and classifying difference equations. By requiring the existence of a symmetry of order $\ell$ and $k$ greater than one, Habibullin et.al.  complemented the conditions contained in Theorem 2 of \cite{ly09} by further integrability conditions. This set of conditions, contained in Proposition 1 of \cite{GH},  reads as follows using our notation:
 
{\bf Proposition 1}. {\sl If an equation of the form (\ref{a4}) admits a higher symmetry of sufficiently great order then the following conditions must hold for some entire $\ell$ and $\ell'$:\\
\noindent
1) $(T_1^{-1}-T_1^{\ell-1})\log f^{(1,1)}_{u_{1,0}}\in Im (I-T_2)$ for $\ell\geq 1$ and $\ell'\leq-1$ ;\\
2) $(T_1-T_1^{\ell'+1})\log f_{u_{1,0}}^{-1,1}\in Im (I-T_2)$ for $\ell\geq 1$ and $\ell'\leq-1$; \\
3) $R_{0,0}( u_{0,1},u_{\ell+1,0},u_{\ell,0},...u_{\ell',0})\in Im (rI-T_2)$ for $\ell\geq 2$ and $\ell'\leq-2$;\\
4) $\widetilde R_{0,0}(u_{0,1},u_{\ell+1,0},u_{\ell,0},...u_{\ell',0})\in Im (\widetilde rI-T_2)$ for $\ell\geq 2$ and $\ell'\leq-2,$
where $I$ is the identity operator. }$$\\ $$
Conditions $1)$ and $2)$ are already contained in \cite{ly09} while $3)$ and $4)$ are higher order and   
$$R_{0,0}=\frac{1}{T_1^{\ell-2}(f^{(1,1)}_{u_{1,0}})} \cdot (T_1^{-1}\left\{f^{(1,1)}_{ u_{0,1}}T_2(g_{u_{\ell,0}})T_1^{\ell-1}(f^{(1,1)}_{u_{1,0}})+f^{(1,1)}_{u{0,0}}g_{u_{\ell,0}}-\right .$$ $$\left .- T_1 T_2(g_{u_{\ell,0}})T_1^{\ell}(f^{(1,1)}_{u_{0,0}})\right\}-T_2(g_{u_{\ell,0}})T_1^{\ell-1}(f^{(1,1)}_{u_{0,1}}).$$
$$r=\frac{T_1^{-1}(f^{(1,1)}_{u_{1,0}})}{T_1^{\ell-2}(f^{(1,1)}_{u_{1,0}})}, \quad \widetilde r=\frac{T_1(f^{-1,1}_{u_{-1,0}})}{T_1^{\ell'+2}(f^{-1,1}_{u_{-1,0}})}$$
$$\widetilde R_{0,0}=\frac{1}{T_1^{\ell'+2}(f^{-1,1}_{u_{-1,0}})} \cdot T_1\left\{f^{-1,1}_{u_{0,1}} T_2 (g_{u_{\ell',0}})T_1^{\ell'+1}(f^{-1,1}_{u_{-1,0}})+f^{-1,1}_{u_{0,0}} g_{u_{\ell',0}}-\right. $$ $$ \left . -T_1^{-1}T_2(g_{u_{\ell',0}})T_1^{\ell
'}(f^{-1,1}_{u_{0,0}})\right\}-T_2(g_{u_{\ell',0}})T_1^{\ell'+1}(f^{-1,1}_{u_{0,1}})\,.$$ 
Conditions  $1)$, $2)$, $3)$ and $4)$ of Proposition 1 provide a sequence of integrability conditions for a given nonlinear partial difference equation in the form of functional equations. Extending the approach used above, which reduced the functional equation to a system of differential equations for the unknown symmetry function $\Phi$, we can derive differential consequences of these equations in the form of a system of first order partial differential equations, which can be solved on characteristics.
 Let us remember, moreover, that if the starting equation is to be integrable then $\ell'=-\ell$. 

So the algorithm for  obtaining the symmetries is similar to the one introduced   by Levi and Yamilov for the case of $3$ point symmetries in the $n$ and $m$ directions, but the operators $A$ and $B$ are  replaced by the characteristic vector fields $Y_k=T_2^{-k} \frac{\partial}{\partial u_{0,1}} T_2^k$, $Y_{-k}=T_2^{k} \frac{\partial}{\partial u_{0,-1}}T_2^{-k}$, $k \in \mathbb Z^+$, introduced by Habibullin et. al. in \cite{GH} and references therein. $Y_1$ and $Y_{-1}$ read: 
\bea \nonumber
 Y_1&=&\frac{\partial}{\partial u_{0,0}}+x_{0,0}\frac{\partial}{\partial  u_{1,0}}+\frac{1}{x_{-1,0}}\frac{\partial}{\partial  u_{-1,0}}+
x_{0,0}x_{1,0}\frac{\partial}{\partial  u_{2,0}}+ \\ \label{defY1} &&\qquad \qquad +\frac{1}{x_{-1,0}x_{-2,0}}\frac{\partial}{\partial u_{-2,0}}+ ...,
\\ \nonumber
 Y_{-1}&=&\frac{\partial}{\partial u_{0,0}}+y_{0,0}\frac{\partial}{\partial  u_{1,0}}+\frac{1}{y_{-1,0}}\frac{\partial}{\partial  u_{-1,0}}+
y_{0,0}y_{1,0}\frac{\partial}{\partial  u_{2,0}}+\\ \label{defY-1} &&\qquad \qquad +\frac{1}{y_{-1,0}y_{-2,0}}\frac{\partial}{\partial u_{-2,0}}+ ...,
\eea
where $$x_{0,0}=T_2^{-1}(\frac{\partial f^{(1,1)}(u_{0,0},u_{1,0},u_{0,1})}{\partial u_{0,1}})=-\frac{\frac{\partial f^{1,-1}(u_{0,0},u_{1,0},u_{0,-1})}{\partial u_{0,0}}}{\frac{\partial f^{1,-1}(u_{0,0},u_{1,0},u_{0,-1})}{\partial u_{1,0}}}$$ and $$y_{0,0}= T_2(\frac{\partial f^{1,-1}(u_{0,0},u_{1,0},u_{0,-1})}{\partial u_{0,-1}})=-\frac{\frac{\partial f(u_{0,0},u_{1,0},u_{0,1})}{\partial u_{0,0}}}{\frac{\partial f(u_{0,0},u_{1,0},u_{0,1})}{\partial u_{1,0}}}.$$ The characteristic vector fields, 
$Y_1$ and $Y_{-1}$,  reduce to $A$ and $B$ when acting on a function that depends only on $u_{n,m}$, $u_{n+1,m}$ and $u_{n-1,m}$. 

If the result of the test, as stated above for a 5 point symmetry, with $3$ points in the $n$ direction and $3$ in the $m$ direction,  is negative,  we have to proceed tentatively. We must look for higher order symmetries following Proposition 1, i.e.  look for a  5-point symmetry  in the $n$ direction and a 5-point symmetry in the $m$ direction, i.e. a $9$  point symmetry, using the operators $Y_1$, $Y_{-1}$, its differential consequences obtained by  applying its commutators and, if necessary, its higher order expression $Y_2$ and $Y_{-2}$, etc..

\section{Integrability properties of (\ref{kj1}).} \label{s2}
In this Section we use the results presented in the previous Section to find the generalized symmetries for (\ref{kj1}), thus proving its integrability. We then study its transformation properties and the relation to other known integrable equations on the square and its relation to (\ref{jk3}).
 
Eq. (\ref{kj1}) satisfies the two necessary conditions in the $n$-direction given in \cite{LY2} but  does not admit a three-points generalized symmetry, either autonomous or not, while  the necessary conditions in the $m$-direction given in \cite{LY2} are not satisfied.
Eq. (\ref{jk3}) is the sum of (\ref{kj1}), (\ref{jk2}) and an arbitrary constant and does not satisfy the integrability conditions given in \cite{LY2} for three-point generalized symmetries either autonomous or not, in either the $n$ or $m$-direction.

So to find the generalized symmetries for (\ref{kj1}) we need to consider higher order symmetries. Applying the theory sketched in the previous Section we find,  after a long and tedious but trivial calculation,  that (\ref{kj1}) has two 5 point symmetries in the $n$ and $m$ direction given by:
\begin{subequations}
\bea
w_{0,0,\epsilon}=w_{0,0}\left(w_{-1,0}w_{0,0}-1\right)\left(w_{0,0}v_{1,0}-1\right)\left(w_{2,0}w_{1,0}-w_{-1,0}w_{-2,0}\right),\label{Carandini2}\\
w_{0,0,\tilde \epsilon}=\frac{w_{0,0}\left(w_{0,-1}+w_{0,0}\right)\left(w_{0,0}+w_{0,1}\right)\left(w_{0,2}+w_{0,1}-w_{0,-1}-w_{0,-2}\right)}{\left(w_{0,-2}+w_{0,-1}+w_{0,0}\right)\left(w_{0,-1}+w_{0,0}+w_{0,1}\right)\left(w_{0,0}+w_{0,1}+w_{0,2}\right)},\label{Carandini3}
\eea
\end{subequations}
where $\epsilon$ and $\tilde \epsilon$ are group parameters.
We notice that  (\ref{kj1}) and both its generalized symmetries are invariant under the discrete transformation $\tilde w_{0,0}\doteq -w_{0,0}$.  The first symmetry  (\ref{Carandini2}) is polynomial and of the form of a Bogoyavlenskyi lattice but of  higher polynomial order. 
The second symmetry  (\ref{Carandini3}) is given by a rational expression where the sum of the order of the numerator and the denominator is equal to the polynomial order of  (\ref{Carandini2}). 

\noindent Analyzing the structure of  (\ref{Carandini2}) we easily see that to reduce its polynomial order  we have to introduce a new  dependent variable $t_{0,0}$ in term of the variable $w_{0,0}$,
\bea
t_{0,0}\doteq w_{0,0}w_{1,0}-1.\label{Carandini4}
\eea
The relation between $t$ and $w$ corresponds to a potentiation as,  by going over to the variables $\tau_{n,m}\doteq-\log ~(t_{n,m}+1)$ and $v_{n,m}\doteq(-1)^n \, \log w_{n,m}$, (\ref{Carandini4}) can be rewritten as $ \tau_{n,m}=v_{n+1,m}-v_{n,m}$. We  thus  can easily invert (\ref{Carandini4})  and we obtain $$w_{n,m}=\left \{v_0 \,e^{- \sum_j \log(t_{j,m}+1)}\right \}^{(-1)^n}.$$

By applying this transformation to the generalized symmetry (\ref{Carandini2}) and (\ref{Carandini3}) we  find the 5-point generalized symmetries of (\ref{kj1})  in the $n$ and $m$ directions,  respectively,  written in the variable $t$:
\bea
 &&t_{0,0,\epsilon}=t_{0,0}\left(t_{0,0}+1\right)\left(t_{2,0}t_{1,0}-t_{-1,0}t_{-2,0}\right),\label{Carandini7}
\\
 &&t_{0,0,\tilde \epsilon}=-\frac{t_{0,0}\left(t_{0,0}+1\right)\left(t_{0,-1}+t_{0,0}+1\right)\left(t_{0,1}+t_{0,0}+1\right)\cdot }{\left[t_{0,2}t_{0,1}+\left(t_{0,0}+1\right)\left(t_{0,2}+t_{0,1}+1\right)\right]\cdot} \\ \nonumber  &&\qquad  \frac{\cdot \left[t_{0,2}t_{0,1}\left(t_{0,-2}+t_{0,-1}+1\right)-t_{0,-2}t_{0,-1}\left(t_{0,2}+t_{0,1}+1\right)\right]}{\cdot \left[t_{0,1}t_{0,-1}  +\left(t_{0,0}+1\right)\left(t_{0,1}+t_{0,-1}+1\right)\right]\left[t_{0,-2}t_{0,-1}+\left(t_{0,0}+1\right)\left(t_{0,-2}+t_{0,-1}+1\right)\right]}.
\eea
Analyzing the structure of (\ref{Carandini3}) we see that we can also transform it to  a Bogoyavlenskyi lattice. In fact,  defining
\bea \label{C1}
\theta_{n,m}&=&\frac{w_{n,m+1}}{w_{n,m}}, \\ \label{C2}
v_{n,m}&=&1+\theta_{n,m}(\theta_{n,m+1}+1), \\ \label{C3}
\tilde t_{n,m}&=&\frac{1}{v_{n,m}},
\eea
i.e. by a potential transformation (\ref{C1}), a Miura transformation (\ref{C2}) and a point transformation (\ref{C3}),  we get the transformation 
\bea \label{C3a}
\tilde t_{n,m}=-\frac{w_{n,m+2}+w_{n,m+1}}{w_{n,m}+w_{n,m+2}+w_{n,m+1}},
\eea
 which transforms (\ref{Carandini3}) into the following Bogoyavlenskyi lattice equation
\bea \label{C4} 
\tilde t_{0,0,\tilde \epsilon}=\tilde t_{0,0}\left(\tilde t_{0,0}+1\right)\left(\tilde t_{0,2}\tilde t_{0,1}-\tilde t_{0,-1}\tilde t_{0,-2}\right).
\eea

 Applying the change of variable  (\ref{Carandini4}) we can transform (\ref{kj1}) to an integrable equation in the variable $t_{0,0}$, given in the following theorem:
\begin{theorem} \label{tt1}
\noindent If (\ref{kj1}) is satisfied, using the definition (\ref{Carandini4}), we have:
\end{theorem}
\begin{subequations}
\bea
&& w_{1,0}=\frac{t_{0,0}+1}{v_{0,0}},\label{Saturnus}\\
&& w_{0,1}=-\frac{t_{0,1}+t_{0,0}+1}{t_{0,0}+1}v_{0,0},\label{Carandini5}\\
&& \quad \left(t_{1,1}+t_{1,0}+1\right)\left(t_{0,1}+t_{0,0}+1\right)-\left(t_{1,0}+1\right)\left(t_{0,1}+1\right)=0.\label{Carandini6}
\eea
\end{subequations}
\noindent{\bf Proof.} To prove (\ref{Saturnus}) solve (\ref{Carandini4}) for $w_{1,0}$; to prove (\ref{Carandini5}) it is sufficient to substitute (\ref{Saturnus}) and its difference consequence,  obtained by shifting once in the $m$ direction,  into  (\ref{kj1}) and solve for $w_{0,1}$; to prove (\ref{Carandini6}) just substitute the expressions for $t_{0,0}$ and all its shifts as given by (\ref{Carandini4}) into its left hand  side, then impose (\ref{kj1}) and its shift obtained by shifting once in the $n$ direction.\qed

Theorem \ref{tt1} shows that, through the potentiation (\ref{Carandini4}), we can transform (\ref{kj1}) into (\ref{Carandini6}), which is an inhomogeneous polynomial equation on the square. Eq. (\ref{Carandini5}) is also a potentiation relation, however in the $m$ direction, which can be solved for either $w$ or $t$ in terms of a summation of a function of the other variable.

\noindent Thus the two relations (\ref{Saturnus}, \ref{Carandini5}) constitute a \emph{potential transformation} between  (\ref{kj1}) and  (\ref{Carandini6}): the compatibility between the $w-$variables implies eq. (\ref{Carandini6}) while the compatibility between the $t-$variables implies  (\ref{kj1}).

 Applying the change of variable  (\ref{C3a}),  we can transform (\ref{kj1}) to an integrable equation in the variable $\tilde t_{n,m}$, as stated in the following theorem:
\begin{theorem}
\noindent If eq. (\ref{kj1}) is satisfied, given (\ref{C3a}),   then
\begin{subequations}
\bea
&&\qquad w_{0,1}=-\frac{\tilde t_{1,0}+\tilde t_{0,0}+1}{\tilde t_{0,0}+1}w_{0,0},\label{Amor1}\\
&&\qquad w_{0,2}=\frac{\tilde t_{1,0}+1}{\tilde t_{0,0}+1}w_{0,0},\label{Amor2}\\
&& \qquad w_{1,0}=\frac{\left(\tilde t_{0,0}+1\right)\left(\tilde t_{1,0}+1\right)}{\left[\tilde t_{2,0}\tilde t_{1,0}+\left(\tilde t_{0,0}+1\right)\left(\tilde t_{2,0}+\tilde t_{1,0}+1\right)\right]w_{0,0}},\label{Amor3}\\
&&\left(\tilde t_{1,1}+\tilde t_{0,1}+1\right)\left(\tilde t_{1,0}+\tilde t_{0,0}+1\right)-\left(\tilde t_{1,0}+1\right)\left(\tilde t_{0,1}+1\right)=0.\label{Romulus4}
\eea
\end{subequations} 
\end{theorem}
\noindent{\bf Proof.} Eq. (\ref{Amor1}) is obtained by solving (\ref{C3a}) for $w_{0,2}$ and substituting the obtained expression, as well as its shift in the $n$ direction, into (\ref{kj1}) shifted once in the $m$ direction, then solving the resulting expression for $w_{1,1}$. Then we substitute the  resulting expression into (\ref{kj1}) and solve for $w_{0,1}$. 

Eq. (\ref{Amor2}) is obtained by substituting (\ref{Amor1}) into the expression found previously for $w_{0,2}$. 

Eq. (\ref{Amor3}) is obtained by substituting (\ref{Amor1}) into the expression found previously for $w_{1,1}$ and  solving  the compatibility condition between this expression and (\ref{Amor1}) for $w_{1,0}$. 

The proof that $\tilde t_{0,0}$ satisfies (\ref{Romulus4}) is obtained by substituting the expressions for $\tilde t_{0,0}$ given by (\ref{C3a}), and any required shifts, into the left hand side of (\ref{Romulus4}), then imposing (\ref{kj1}) as well as its first and second shifts in the $m$ direction. \qed

\noindent The three relations (\ref{Amor1}, \ref{Amor2}, \ref{Amor3}) constitute a \emph{Miura transformation} between  (\ref{kj1}) and  (\ref{Romulus4}). The compatibility between the $v-$variables implies  (\ref{Romulus4}) while the compatibility between the $\tilde t-$variables implies  (\ref{kj1}).

From  (\ref{Carandini6}) we can infer the integrability of two non-trivial,  non-autonomous extensions of (\ref{kj1}). The first extension is provided by the following theorem: 
\begin{theorem}
\noindent If  (\ref{Carandini6}) is satisfied, given (\ref{Saturnus}) as a definition for the variable $w_{0,0}$, then $w_{0,0}$ satisfies the non autonomous equation
\bea
w_{0,0}w_{1,0}+w_{1,0}w_{0,1}\left(1-\frac{2}{\gamma_{m}\left(-1\right)^n+1}\right)+w_{0,1}w_{1,1}-1=0,\label{Carandini8}
\eea
where $\gamma_{m}$ is an arbitrary function of $m$. The solution $\tilde w_{0,0}$ of (\ref{Carandini8})    and the solution $w_{0,0}$  of (\ref{kj1}) are related by the following point wise transformation:
\bea \label{Romulus1}
\tilde w_{0,0}\doteq\kappa_{m}^{\left(-1\right)^{n+1}}w_{0,0},\ \ \ \kappa_{m}\not=0,\\  \kappa_{m+1}=\frac{\gamma_{m}+1}{\gamma_{m}-1}\kappa_{m},\label{Romulus1a}
\eea
 where $\kappa_{m}$ is an arbitrary function of $m$, solution of (\ref{Romulus1a}).
\end{theorem}
\noindent{\bf Proof.} Substitution of (\ref{Saturnus}) and its shifts into (\ref{Carandini6}), gives the equation
\bea \label{Carandini8a}
\alpha_{0,0}+\alpha_{1,0}-\alpha_{0,0}\alpha_{1,0}=0,\ \ \ \alpha_{0,0}\doteq\frac{\mathcal{E}_{0,0}}{w_{1,0}w_{0,1}},
\eea
which, if $\alpha_{0,0}=0$, implies $\mathcal{E}_{0,0}=w_{0,0}w_{1,0}+w_{1,0}w_{0,1}+w_{0,1}w_{1,1}-1=0$, which is nothing  but equation (\ref{kj1}). If $\alpha_{0,0}\not=0$ and we introduce the variable $z_{0,0}\doteq 1/\alpha_{0,0}$, (\ref{Carandini8a}) linearizes to $z_{1,0}+z_{0,0}-1=0$, which gives  (\ref{Carandini8}) when integrated. One could prove that the converse is also true by straightforward calculation.

\noindent As relation (\ref{Saturnus}) is covariant under (\ref{Romulus1}), given (\ref{Saturnus}, \ref{Carandini6}), we can always \emph{fix a gauge} for $w_{0,0}$ so that (\ref{kj1}) is satisfied.\qed
 
 The second extension is provided by the following theorem:
\begin{theorem}
\noindent If  (\ref{Carandini6}) is satisfied, then the variable $w_{0,0}$ defined by (\ref{Carandini5}) satisfies the non autonomous equation
\bea
w_{0,0}w_{1,0}+w_{1,0}w_{0,1}+w_{0,1}w_{1,1}-(1+\beta_{n})=0,\ \ \ \beta_{n}\not=-1,\label{Carandini9}
\eea
where $\beta_{n}$ is an arbitrary function of $n$.

The solution $\tilde w_{0,0}$ of (\ref{Carandini9})    and the solution $w_{0,0}$  of (\ref{kj1}) are related by the following point wise transformation:
\bea
w_{0,0}^{\prime}\doteq\epsilon_{n}^{-1}w_{0,0},\ \ \ \epsilon_{n}\not=0,\ \ \ \beta_{n}=\epsilon_{n+1}\epsilon_{n}-1,\label{Romulus2}
\eea
where $\epsilon_{n}$ is an arbitrary function of $n$.
\end{theorem}
\noindent{\bf Proof.} Inserting into (\ref{Carandini6}) the expression for $t_{0,1}$ derived from (\ref{Carandini5}) and its shift in the $n$ direction, we obtain
\bea
t_{0,0}=-\frac{\theta_{0,0}\left(\theta_{1,0}+1\right)}{\theta_{0,0}\left(\theta_{1,0}+1\right)+1},\ \ \ \theta_{0,0}\doteq\frac{w_{0,1}}{v_{0,0}},\label{kj10}
\eea
\noindent where  $\theta_{0,0}\left(\theta_{1,0}+1\right)+1\not=0$ because $\mathcal{E}_{0,0}\not=-1$. The compatibility between (\ref{kj10}) and (\ref{Carandini5}) gives
\bea
1+\theta_{0,0}-\theta_{0,0}\theta_{1,0}\theta_{0,1}\left(1+\theta_{1,1}\right)=0\Rightarrow\mathcal{E}_{0,0}=\mathcal{E}_{0,1},
\eea
\noindent which, when integrated, gives eq. (\ref{Carandini9}) (the condition $\beta_{n}\not=-1$ follows directly from $\mathcal{E}_{0,0}\not=-1$).

\noindent As relation (\ref{Carandini5}) is covariant under (\ref{Romulus2}), given (\ref{Carandini5}, \ref{Carandini6}), we can always \emph{fix a gauge} for $w_{0,0}$ so that (\ref{kj1}) is satisfied.\qed

In  \cite{MX} the authors present the nonlinear partial difference equation on the square 
\bea
\left(u_{0,0}+u_{1,1}\right)u_{1,0}u_{0,1}+1=0.\label{kj11}
\eea
whose lowest generalized symmetry involves  just 5 points in each direction and is given by the Bogoyavlenskyi lattice  (\ref{Carandini7}) in the variable $t_{0,0}$,  defined by
\bea
t_{0,0}\doteq\frac{1}{u_{-1,0}u_{0,0}u_{1,0}-1}.\label{kj12}
\eea
This leads to the  following theorem:
\begin{theorem}
\noindent If (\ref{kj11}) is satisfied, given (\ref{kj12}), then
\begin{subequations}
\bea
u_{1,0}&=&\frac{t_{0,0}+1}{t_{0,0}u_{-1,0}u_{0,0}},\label{Amaryllis}\\
u_{0,1}&=&-\frac{t_{0,1}+t_{0,0}+1}{t_{0,0}+1}u_{-1,0},\label{kj13}\\
u_{-1,1}&=&\frac{t_{0,0}+1}{t_{0,1}u_{-1,0}u_{0,0}},\label{kj14}
\eea
\end{subequations}
and $t_{0,0}$ satisfies  (\ref{Carandini6}).
\end{theorem}
\noindent{\bf Proof.} Eq. (\ref{Amaryllis}) follows directly from (\ref{kj12}). 

\noindent To prove (\ref{kj13}) one has to: a) substitute (\ref{Amaryllis}) and its $m$-shift into (\ref{kj11}) and solve for $u_{-1,1}$; b) substitute the resulting expression into (\ref{kj11}) back-shifted once in the $n$-direction. 

\noindent To prove (\ref{kj14}) just substitute (\ref{kj13}) into the previous expression for $u_{-1,1}$. 

\noindent To prove that $t_{0,0}$ satisfies  (\ref{Carandini6}), substitute (\ref{kj12}) and any required shifts into the left hand side of (\ref{Carandini6}), then impose (\ref{kj11}) as well as its shift and back-shift in the $n$-direction. \qed

\noindent The three relations (\ref{Amaryllis}, \ref{kj13}, \ref{kj14}) constitute a \emph{potential transformation} between (\ref{Carandini6}) (for $t_{0,0}$) and (\ref{kj11}): the compatibility between the $t-$variables implies  (\ref{kj11}) while the compatibility between the $u-$variables implies (\ref{Carandini6}) (for $t_{0,0}^{\prime}$).

When $t_{0,0}\not=0$, we can define the variable
\bea
z_{0,0}\doteq\frac{1}{t_{0,0}}.\label{kj15}
\eea
and we have the following theorem which is proved by straightforward calculation:
\begin{theorem}
\noindent If  (\ref{Carandini6}) is satisfied, given (\ref{kj15}), then
\bea
z_{0,0}z_{1,0}+z_{0,1}z_{1,1}+z_{1,0}z_{0,1}\left(z_{1,1}+z_{0,0}+1\right)=0.\label{kj16}
\eea
\end{theorem}
Eq. (\ref{kj16}) is  (\ref{jk3}) when $\chi=0$.  The calculation of the symmetries of  (\ref{jk3}) turn out to be too complicate even though it should be strictly related  by the transformation (\ref{kj15}) to the Bogoyavlenskyi lattice, the symmetry of (\ref{Carandini6}). So to show its integrability we rely on the calculation of its algebraic entropy.  
\subsection{Algebraic entropy}
Eq. (\ref{jk3}) is an equation for the
 dependent variable $w_{n,m}$ that depends on two discrete independent variables, $n$ and $m$, and its shifts. The entropy is zero for this equation and the growth of degrees is quadratic, indicating that the equation is integrable and not linearizable. This was known for the case where the parameter was set to $\chi=0$, but it also occurs in the general case for arbitrary $\chi$.

To be self-contained, we include the following brief description of algebraic entropy based on \cite{v06}. Finding the algebraic entropy of an equation requires us to begin with an initial condition and  then use the equation to find values of the field over the other points on the lattice. The initial condition is usually chosen to be a sequence of arbitrary values along a staircase of points through the lattice. Because of the form of the equation, the field values will always be rational functions of the initial conditions. The degree at any point, denoted $d^{(\ell)}$, is the greatest polynomial degree of either the numerator or denominator of the field value, after any possible cancellations have been performed. Only one discrete index, $\ell$, is needed, this represents the distance of the point in question from the staircase of initial conditions. The algebraic entropy is then given by:
$$
\epsilon=\lim_{\ell\rightarrow \infty}\frac{1}{\ell}\log{d^{(\ell)}}. 
$$

There are four fundamental directions of evolution related to the direction chosen for the staircase of initial conditions (see Section 4 of \cite{v06}). Thus there are four sequences of degrees for an equation on a lattice in general, and in this case we find the following pairs of degree sequences:
$$
d^{(\ell)}_{-+}=d^{(\ell)}_{+-}=\{1,3,7,13,21,31,43,57,73,91,111,\ldots\},
$$
and
$$
d^{(\ell)}_{++}=d^{(\ell)}_{--}=\{1,2,5,10,16,24,35,46,59,76,92,110,133,154,177,206\ldots\},
$$
where the signs in the subscript indicate the direction of the associated staircase of initial conditions.

Although it is usually impossible to calculate an infinite sequence of degrees and use the definition of algebraic entropy directly, if a generating function can be found for the sequence of degrees, we can use it to calculated the entropy. Obtaining generating functions for the above sequences, we find
$$
g_{-+}(t)=g_{+-}(t)=\frac{1+t^2}{(1-t)^3},
$$
and
$$
g_{++}(t)=g_{--}(t)=\frac{1+t+3 t^2+3 t^3+4 t^4+2 t^5+2 t^6}{(1-t)^3 (1+t+t^2)^2}.
$$
Generating functions are closely related to the $Z$-transform, which is a discrete Laplace transform. The variable $t$, in $g(t)$, is the transform variable. The entropy is given by the logarithm of the inverse of the modulus of the smallest pole of the generating function. This shows that the entropy vanishes, the growth of degrees is quadratic in each direction and the equation (\ref{jk3}) should be integrable.
\subsection{ The Lax pair representation.}
\noindent Let us consider the Lax pair representation of  (\ref{kj11}) \cite{MX} given by
\bea
&&  \Psi_{1,0}=M_{0,0}\Psi_{0,0},\ \ \ \Psi_{0,1}=N_{0,0}\Psi_{0,0},\nonumber\\
&&M_{0,0}\doteq\left(\begin{array}{ccc}
0 & 1 & 0\\
-u_{0,0} & -u_{0,0}u_{1,0} & \lambda\\
-1 & 0 & \frac{1}{u_{0,0}}\end{array}\right),\, N_{0,0}\doteq\left(\begin{array}{ccc}
0 & 0 & 1\\
-1 & 0 & \frac{1}{u_{0,0}}\\
\frac{u_{0,0}u_{0,1}}{\lambda} & -\frac{1}{\lambda} & 0\end{array}\right),\nonumber
\eea
\noindent and perform the gauge transformation
\bea
\Psi_{0,0}\doteq V_{0,0}\Phi_{1,0},\ \ \ V_{0,0}\doteq\left(\begin{array}{ccc}
1 & 0 & 0\\
0 & u_{0,0}u_{-1,0} & 0\\
0 & 0 & u_{0,0}\end{array}\right),
\eea
\noindent so that
\bea
&&\Phi_{1,0}=\frac{1}{u_{0,0}}\tilde M_{0,0}\Phi_{0,0},\ \ \ \Phi_{0,1}=u_{-1,0}\tilde N_{0,0}\Phi_{0,0},\label{kj17}\\
&&\tilde M_{0,0}\doteq u_{0,0}V_{0,0}^{-1}\cdot M_{-1,0}\cdot V_{-1,0},\ \ \ \tilde N_{0,0}\doteq\frac{1}{u_{-1,0}}V_{-1,1}^{-1}\cdot N_{-1,0}\cdot V_{-1,0}.\nonumber
\eea
\noindent Using (\ref{Carandini6}, \ref{Amaryllis}, \ref{kj13}, \ref{kj14}) (the last three written for $t_{0,0}$) and their shifts, we get
\begin{subequations}
\bea
&&\tilde M_{0,0}=\left(\begin{array}{ccc}
0 & 1+\frac{1}{t_{-1,0}} & 0\\
-1 & -1-\frac{1}{t_{-1,0}} & \lambda\\
-1 & 0 & 1\end{array}\right),\label{kj18}\\
&&\tilde N_{0,0}=\left(\begin{array}{ccc}
0 & 0 & 1\\
1+\frac{t_{0,1}\left(t_{-1,0}+1\right)}{t_{-1,0}\left(t_{0,0}+1\right)} & 0 & -1-\frac{t_{0,1}\left(t_{-1,0}+1\right)}{t_{-1,0}\left(t_{0,0}+1\right)}\\
\frac{1}{\lambda} & -\frac{t_{0,1}\left(t_{-1,0}+1\right)}{t_{-1,0}\left(t_{0,0}+1\right)\lambda} & 0\end{array}\right).\label{kj19}
\eea
\end{subequations}
\noindent The compatibility condition between (\ref{kj17}) reads
\bea \label{kk1}
\tilde M_{0,1}\tilde N_{0,0}=\frac{u_{0,1}}{u_{-1,0}}\tilde N_{1,0}\tilde M_{0,0},
\eea
\noindent so that, removing $u_{0,0}$ from (\ref{kk1}) and using (\ref{kj13}), we have that (\ref{kj17}, \ref{kj18}, \ref{kj19}) is a nonlocal Lax pair representation of eq. (\ref{Carandini6}). Alternatively from (\ref{kj17}) we have
\bea
\Phi_{3,0}=Y_{0,0}\Phi_{0,0},\ \ \ \Phi_{0,1}=u_{-1,0}\tilde N_{0,0}\Phi_{0,0},\ \ \ Y_{0,0}\doteq\frac{1}{u_{2,0}u_{1,0}u_{0,0}}\tilde M_{2,0}\tilde M_{1,0}\tilde M_{0,0},\label{Carandini20}
\eea
\noindent whose compatibility gives
\bea
Y_{0,1}\tilde N_{0,0}=\frac{u_{2,0}}{u_{-1,0}}\tilde N_{3,0}Y_{0,0}.
\eea
Using (\ref{Amaryllis}), which implies that $u_{2,0}-\frac{t_{0,0}\left(t_{1,0}+1\right)}{t_{1,0}\left(t_{0,0}+1\right)}u_{-1,0}=0$, to remove $u_{0,0}$ from the compatibility condition above, we have that (\ref{Carandini20}, \ref{kj18}, \ref{kj19}) is a (local in the $m$-direction) Lax pair representation of  (\ref{Carandini6}). Finally from (\ref{Carandini20}) we have
\bea
\Phi_{3,0}=Y_{0,0}\Phi_{0,0},\ \ \ \Phi_{0,3}=X_{0,0}\Phi_{0,0},\ \ \ X_{0,0}\doteq u_{-1,2}u_{-1,1}u_{-1,0}\tilde N_{0,2}\tilde N_{0,1}\tilde N_{0,0}.\label{Carandini21}
\eea
Using (\ref{kj13}) and (\ref{kj14}), which imply $u_{-1,2}u_{-1,1}u_{-1,0}=-\frac{\left(t_{0,1}+1\right)\left(t_{0,0}+1\right)}{t_{0,2}\left(t_{0,1}+t_{0,0}+1\right)}$, we have that (\ref{Carandini21}, \ref{kj18}, \ref{kj19}) is a local Lax pair representation of eq. (\ref{Carandini6}). Using (\ref{kj1}, \ref{Saturnus}, \ref{Carandini5}) and their shifts, we obtain the corresponding Lax pair representations of eq. (\ref{kj1}) while, using (\ref{kj15}), we obtain the corresponding Lax pair representations of eq. (\ref{kj16}).

\section{Conclusions} \label{s3}
In this article we have shown that the equation (\ref{kj1}), obtained from the multiple scale expansion of a quite general class of dispersive multilinear equation on a four point lattice at the order $\epsilon^6$, is integrable. We do so by calculating its generalized symmetries and by finding its Lax pair. In doing so we are able to show that (\ref{kj1}) is related by potentiation or Miura transformation to the integrable equation  (\ref{kj11}), introduced recently by Mikhailov and Xenitidis in \cite{MX} where they show its integrability, and to, up to our knowledge, the new equation (\ref{Carandini6}). Moreover, from the connection between (\ref{kj1}) and (\ref{Carandini6}), we are able to show that two non autonomous  extensions of (\ref{kj1}), (\ref{Carandini8}) and (\ref{Carandini9}) are also integrable. 

The proof of the existence of generalized symmetries for the equation (\ref{jk3})
when $\chi\not=0$ is left for future work as its integrability is shown
only by algebraic entropy. The integrability of the other equations
obtained by the $\epsilon^6$ multiple scale expansion, (\ref{Quadrelli2}, \ref{Quadrelli3}) when
$\tau\not=0$, will also be left for future consideration.

\section*{Acknowledgments}
DL and CS have been partly supported by the Italian Ministry of Education and Research, 2010 PRIN ÒContinuous and discrete nonlinear integrable evolutions: from water waves to symplectic mapsÓ. We thank R.I. Yamilov and J. Hietarinta for many enlightening discussions.

\end{document}